# Scenario Generation for Cooling, Heating, and Power Loads Using Generative Moment Matching Networks

Wenlong Liao, Yusen Wang, Yuelong Wang, Kody Powell, Qi Liu, and Zhe Yang

*Abstract*—Scenario generations of cooling, heating, and power loads are of great significance for the economic operation and stability analysis of integrated energy systems. In this paper, a novel deep generative network is proposed to model cooling, heating, and power load curves based on a generative moment matching networks (GMMN) where an auto-encoder transforms high-dimensional load curves into low-dimensional latent variables and the maximum mean discrepancy represents the similarity metrics between the generated samples and the real samples. After training the model, the new scenarios are generated by feeding Gaussian noises to the scenario generator of the GMMN. Unlike the explicit density models, the proposed GMMN does not need to artificially assume the probability distribution of the load curves, which leads to stronger universality. The simulation results show that the GMMN not only fits the probability distribution of multi-class load curves well, but also accurately captures the shape (e.g., large peaks, fast ramps, and fluctuation), frequency-domain characteristics, and temporal-spatial correlations of cooling, heating, and power loads. Furthermore, the energy consumption of generated samples closely resembles that of real samples.

## NOMENCLATURE

*Abbreviations*

| | |
|---|---|
| HMM | hidden Markov model |
| GAN | generative adversarial network |
| VAE | variational auto-encoder |
| GMMN | generative moment matching network |
| MMD | maximum mean discrepancy |
| MSE | mean square error |
| ReLU | rectified linear unit |
| PSD | power spectral density |
| PDF | probability distribution function |
| LSTM | long short-term memory |
| GRU | the gated recurrent unit |

*Parameters*

| | |
|---|---|
| $X_{e,i}$ | a historical power load curve |
| $X_{c,i}$ | a historical cooling load curve |
| $X_{h,i}$ | a historical heating load curve |
| $X_i$ | an input sample of a GMMN |
| $X_g$ | new load curves generated by the scenario generator |
| $H$ | the low-dimensional latent variables |
| $W_g, W_e, W_d$ | weight matrixes of different models |
| $B_g, B_e, B_d$ | bias vectors of different models |
| $f_g, f_e, f_d$ | activation functions of different models |
| $Z$ | a noise vector |
| $N$ | the number of generated load curves at each iteration |
| $M$ | the number of real load curves at each iteration |
| $\phi$ | a transformation function |
| $k(\cdot)$ | a kind of kernel |
| $x'_{g,i}$ | the representation of generated load curves in the new space |
| $x'_{r,i}$ | the representation of real load curves in the new space |
| $\upsilon$ | the bandwidth parameter |
| $L_{AE}$ | the loss function of an auto-encoder |
| $\mathcal{L}_{MMD^2}$ | the loss function of a scenario generator |

W. Liao and Z. Yang are with the Department of Energy Technology, Aalborg University, Aalborg 9220, Denmark.
Y. Wang is with the School of Electrical Engineering and Computer Science, KTH Royal Institute of Technology, Stockholm, SE-100 44, Sweden.
Y. Wang is with the State Grid Tianjin Chengxi Electric Power Supply Branch, Tianjin 300100, China.
K. Powell is with the Department of Chemical Engineering, University of Utah, UT 84112, America.
Q. Liu (Corresponding author, e-mail: liuqi_tj@tju.edu.cn) is with the Key Laboratory of Smart Grid of Ministry of Education, Tianjin University, Tianjin 300072, China.

## I. INTRODUCTION

Integrated energy systems coupled with cooling, heating, and electric power energies can improve energy efficiency and meet the needs of islands, which have become increasingly popular in recent years [1]. To better coordinate and control flexible resources in integrated energy systems such as heat pumps, electric vehicles, stored energy, and air conditioners, it is necessary to accurately model cooling, heating, and power loads [2]. One widely used method to model these loads in integrated energy systems is generating a set of stochastic scenarios. By using a set of possible time series scenarios, system operators can make decisions which account for the uncertainties of cooling, heating, and power loads, such as stochastic optimization and robust optimization [3]. Therefore, the scenario generations of cooling, heating, and power loads are of great significance for the operation and planning of integrated energy systems.

The main idea of stochastic scenario generation is to generate a set of new samples similar to the historical load curves, which are used to train generative models. With respect to whether the probability distributions of load curves are needed, the existing methods for stochastic scenario generation can be divided into two categories: explicit density models and implicit density models [4]. Specifically, explicit density models need to artificially assume the probability distribution of load curves, and use historical samples to fit the key parameters in the probability distribution. For example, a Gaussian mixture model is proposed to represent the distribution characteristics of power loads in [5], and then the Monte Carlo method is used to generate stochastic scenarios of the power loads. Similarly, Ref. [6] approximated the probability distribution of loads with a normal distribution, and obtained new load curves through the Latin hypercube sampling method. In order to take into account the spatial correlation between multiple nodes when generating stochastic scenarios, the Copula theory is used to construct a joint distribution function of loads in [7], and the load curves of multiple nodes are obtained simultaneously by sampling. In general, load curves generated by explicit density models are of poor quality, since they rely on the probability distribution of the load curves which are assumed artificially. Additionally, the probability distribution of load curves is unknown most of the time, and it is difficult to accurately represent it with mathematical formulas. The probability distributions of load curves in different times and regions are also different, which makes the explicit density models not universal [8].

In contrast, the implicit density models do not require explicit likelihood estimation or artificial assumption of the probability distribution of the load curves. After training models, stochastic scenarios obeying the potential probability distribution are obtained by inputting noises to the scenario generator. In addition, the implicit density models can be applied to stochastic scenario generation of various loads in different times and regions by adjusting the structure and parameters [9]. For stochastic scenario generation in integrated energy systems, existing implicit density models mainly include the hidden Markov model (HMM), generative adversarial network (GAN), and variational auto-encoder (VAE) [10]. Specifically, HMM is often used in data generation tasks, because of its simple structure and clear physical meaning [11]. However, due to the assumption of independence of its output variables, the context information is ignored in HMM, and it is difficult to capture the spatial-temporal characteristics of the load curves. Both VAE and GAN are powerful generative models in deep learning, and have been extensively and independently studied in the stochastic scenario generation tasks of distribution networks [12]. Nevertheless, VAE can only approximate the lower bound of the log-likelihood of the real load curves, which leads to the poor quality of the new samples generated by VAE [13]. The vanishing gradients and exploding gradients problems of the GAN in the training process still exist in previous publications and these problems limit the quality of generated scenarios [14], [15].

The generative moment matching network (GMMN) is a new deep generative network widely used in the field of computer vision [16]. Compared with the other generative networks such as the VAE and the GAN, the GMMN presents the more stable training process and higher quality generated samples [17], since it directly uses the maximum mean discrepancy (MMD) to represent the similarity metrics between the generated samples and the real samples. At present, GMMN has shown excellent performance in many fields such as image de-noising, image generation, voice synthesis, and style transfer [18], [19]. The successful applications of the GMMN in the image and videos prove that it can learn complex objective laws of high-dimensional data through unsupervised training. The application of the GMMN in energy systems is relatively limited. In [20], GMMN is used to model a single wind power curve without considering the correlation among multiple wind farms. In theory, GMMN can not only use deep convolutional neural networks with strong learning ability to effectively extract latent representations from cooling, heating, and power loads, but also employ the MMD as the loss function to reduce the distance between generated samples and real samples, so as to greatly improve the quality of new stochastic scenarios. However, the existing structures and parameters of the GMMN are designed for computer vision, which is not suitable for the 1-dimensional time series of loads. Therefore, it is still a challenge to design a GMMN for scenario generation of cooling, heating, and power loads considering their temporal correlations and spatial correlations. Moreover, How to fine-tune the parameters of the GMMN for scenario generation of multi-class loads, and whether the GMMN shows outstanding performance for scenario generation deserve further study.

In this paper, it is aimed to generalize the GMMN for scenario generation of cooling, heating, and power loads considering their temporal correlation and spatial correlation. The performance of the proposed method is tested by a real-world dataset. The key contributions of this paper include:

1) A novel data-driven method is proposed for stochastic scenario generation of cooling, heating, and electric power loads. By employing the deep convolutional neural network and MMD, it accurately captures the hallmark characteristics (e.g., large peaks, fast ramps, and fluctuation), frequency-domain characteristics, distribution characteristics, and temporal correlations of cooling, heating, and power loads.

2) Unlike the explicit density models, the proposed GMMN does not need to artificially assume the probability distribution of the load curves, which leads to stronger universality. By adjusting the structure and parameters of the network, GMMN can simultaneously generate stochastic scenarios of cooling, heating, and power loads accounting for spatial correlations. After training, Gaussian noises are fed to the GMMN to generate any number of stochastic scenarios, which provides sufficient data support for uncertain optimization and decision-making of integrated energy systems.

3) Extensive experiments on a real dataset collected from the University of Texas at Austin are performed to validate the effectiveness of the GMMN for scenario generations. The influence of key parameters of the GMMN (e.g. the types and numbers of hidden layers, optimizer, and the learning rate) on the performance is analyzed, and the constructive suggestions for how the select these parameters are given.

The rest of this paper is organized as follows. Section II explains the structure and parameters of the GMMN. Section

III introduces the process of stochastic scenario generation based on a GMMN. Section IV performs the simulations. Section V discusses the limitation and generalization of the proposed approach. Section VI summarizes the conclusions.

## II. GENERATIVE MOMENT MATCHING NETWORKS

### A. Problem Formulation for Scenario Generation

Suppose $X_{e,i} = \{x_{e,1}, x_{e,2}, \ldots, x_{e,T}\}$ is a historical power load curve indexed by time, $t=1, \ldots, T$, and $i$ ranges from 1 to $N_e$. The objective is to train a scenario generator by using $N_e$ historical power load curves. The new power load curve generated by the scenario generator should have the same stochastic processes and similar properties as historical power load curves.

There are strong correlations between the cooling, heating, and power loads in integrated energy systems [21]-[23]. Therefore, when scenario generation of power load is generalized into multi-class loads, their correlations need to be considered. Suppose $X_{c,i} = \{x_{c,1}, x_{c,2}, \ldots, x_{c,T}\}$ is a historical cooling load curve where $i$ ranges from 1 to $N_c$, and $X_{h,i} = \{x_{h,1}, x_{h,2}, \ldots, x_{h,T}\}$ is a historical heating load curve where $i$ ranges from 1 to $N_h$. Generated scenarios should be capable of capturing both the temporal correlations and spatial correlations between the cooling, heating, and power loads, as well as the probability distribution of each load.

Note that if three GMMNs are trained independently for cooling, heating, and power loads, their spatial correlation will be lost. To account for the spatial correlations between the loads, a cooling load curve $X_{c,i}$, a heating load curve $X_{h,i}$, and a power load curve $X_{e,i}$ form a total input sample $X_i$ of the GMMN. In this case, the input samples of the GMMN contain information about spatial correlations, and the goal of the GMMN is to make the generated samples and the input samples similar enough, i.e., the correlation will be considered during the training process. In short, only one GMMN needs to be trained, and it is able to simultaneously generate stochastic scenarios for cooling, heating, and power loads considering their correlations [4].

### B. Scenario Generation Using the GMMN

As shown in Fig. 1, a GMMN consists of an auto-encoder and a scenario generator. Firstly, the real samples are used to train the auto-encoder composed of an encoder and a decoder. The mean square error (MSE) between the real samples and the generated fake samples is regarded as the loss function, which is utilized to update the weights of the auto-encoder. Secondly, Gaussian noises are fed to the scenario generator to obtain new samples. To update the weights of the scenario generator, the MMD between the new sample and the real samples is calculated by the trained encoder from the auto-encoder. After the training, the stochastic scenarios of cooling, heating, and power loads can be obtained by feeding Gaussian noises to the scenario generator.

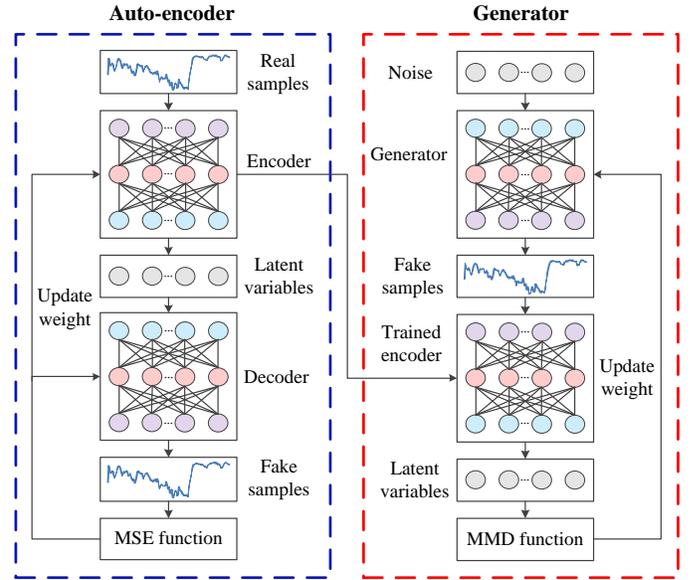

Fig. 1. The framework of the GMMN for stochastic scenario generation.

*1) The loss function and structures of the scenario generator*

The main idea of the scenario generator is to sample a simple prior distribution $Z \sim P_{z(z)}$ (e.g., Gaussian distribution) to obtain a noise vector $Z$. Then, a convolutional neural network composed of transposed convolutional layers is selected to represent the complex nonlinear relationship between the noise vector $Z$ and the real load curves due to its powerful feature extraction ability [24]. The mathematical formula of transposed convolutional layers is:

$$X_g = f_g(Z * W_g + B_g) \quad (1)$$

where $X_g$ denotes the new load curves generated by the scenario generator; $f_g$ denotes the activation function of the scenario generator; $W_g$ and $B_g$ denote the weight matrix and bias vector of the scenario generator, respectively; and $*$ denotes the transposed convolutional operation.

Unlike GAN, whose training process involves difficult min-max optimization problems, GMMN is comparatively simple, since it chooses to minimize a straightforward loss function. Specifically, MMD is a very popular statistical distance to compare the similarity metrics between two datasets and whether the samples are from the same distribution [25]. Therefore, MMD is used to measure the difference between the generated load curves $X_g$ and the real load curves $X_r$. Its mathematical formula is:

$$\mathcal{L}_{\text{MMD}^2} = \left\| \frac{1}{N} \sum_{i=1}^{N} \phi(x_{g,i}) - \frac{1}{M} \sum_{j=1}^{M} \phi(x_{r,j}) \right\|^2 \quad (2)$$

where $N$ denotes the number of generated load curves at each iteration; $M$ denotes the number of real load curves at each iteration; and $\phi$ denotes a transformation function, which leads to matching the difference of sample. Note that $M$ is equal to $N$.

Obviously, each term in Eq. (2) only involves inner products between vectors, and thus load curves can be transferred into a new space by kernel tricks. Its mathematical formula is:

$$\mathcal{L}_{MMD^2} = \frac{1}{N^2}\sum_{i=1}^{N}\sum_{i'=1}^{N}k\left(x_{g,i}, x'_{g,i}\right) - \frac{2}{NM}\sum_{i=1}^{N}\sum_{j=1}^{M}k\left(x_{g,i}, x_{r,j}\right)$$
$$+ \frac{1}{M^2}\sum_{j=1}^{M}\sum_{j'=1}^{M}k\left(x_{g,j}, x'_{r,j}\right) \quad (3)$$

where $k(\cdot)$ denotes a kind of kernel; $x'_{g,i}$ denotes the representation of generated load curves in the new space; and $x'_{r,i}$ is the representation of real load curves in the new space.

Furthermore, if $\phi$ in Eq. (2) is an identity transformation, MMD is equivalent to the mean difference between the generated load curves and the real load curves. In this case, the GMMN can be considered as an auto-encoder. If $\phi$ is a quadratic transformation, MMD is equivalent to the second-order moment between the generated load curves and the real load curves. In the same way, If $\phi$ includes all term transformations, MMD covers all order moments and the probability distribution of load curves [26], so this network structure is called the generative moment matching network.

To make the generated load curves and the real load curves have the same characteristics (e.g., probability distribution, shapes, and correlations), the GMMN should include all term transformation. The Gaussian function can be converted into an infinite series through Taylor expansion, which just meets the requirement of Eq. (2) to calculate each moment. Therefore, the kernel in Eq. (3) uses Gaussian kernel:

$$k(x, x') = \exp\left(-\frac{1}{2\upsilon}|x-x'|^2\right) \quad (4)$$

where $\upsilon$ is the bandwidth parameter. So far, GMMN has obtained the loss function, and the weight of the generative model can be updated by the chain rule and gradient descent method to complete the training process.

2) The loss function and structures of the auto-encoder

Due to the need to generate stochastic scenarios of cooling, heating, and power loads at the same time, the dimension of load curves generated by the GMMN is very high, which is not conducive to calculating the loss function. Fortunately, there are strong temporal-spatial correlations among cooling, heating, and power loads [27], i.e. high-dimensional load curves can be represented by low-dimensional manifold. This is beneficial for statistical estimators such as the MMD, since the volume of data required to generate a reliable estimator grows with the dimension of the data [28]. Therefore, this paper uses an auto-encoder to map the high-dimensional load curves into low-dimensional latent variables, which are used to calculate the MMD loss function.

The auto-encoder is a kind of unsupervised neural network that is composed of an encoder and a decoder, and its loss function is to make the input data equal to the output data through unsupervised learning [29]. Specifically, the encoder maps high-dimensional load curves to low-dimensional latent variables, which reflect the main characteristics of the original input data. Then, the decoder reconstructs the latent variables into new load curves similar to the input data.

Take the encoder and decoder constructed by dense layers as an example to illustrate the data stream transmission process of the auto-encoder. In the encoding process, the input data of the encoder is the real load curves $X_r$, which are passed through multiple dense layers to obtain low-dimensional latent variables $H$. The mathematical formula of the encoder is:

$$H = f_e\left(X_r W_e + B_e\right) \quad (5)$$

where $f_e$ denotes the activation function of the encoder; $W_e$ and $B_e$ denote the weight matrix and bias vector of the encoder, respectively.

In the decoding process, the input data of the decoder is the low-dimensional latent variables $H$, which are passed through multiple dense layers to obtain reconstructed load curves $X_d$. The mathematical formula of the decoder is:

$$X_d = f_d\left(HW_d + B_d\right) \quad (6)$$

where $f_d$ denotes the activation function of the decoder; $W_d$ and $B_d$ denote the weight matrix and bias vector of the decoder, respectively.

The goal of auto-encoder is to make the real load curves and reconstructed load curves as similar as possible, so the loss function $L_{AE}$ can be defined as MSE:

$$L_{AE} = \frac{1}{M}\sum_{i=1}^{M}\left(x_{r,i} - x_{d,i}\right)^2 \quad (7)$$

where $M$ is the number of real load curves at each iteration; $X_{r,i}$ is the $i^{th}$ elements of the real load curves; and $X_{d,i}$ is the $i^{th}$ elements of the reconstructed load curves.

III. STOCHASTIC SCENARIO GENERATION VIA GMMN

When the GMMN is used for stochastic scenario generation of loads, the scenario generator of the GMMN will be affected by physical factors such as the temporal-spatial correlations between cooling, heating, and power loads. Although stochastic scenario generation is different from the data generation in the field of computer vision, the process is similar. Specifically, the core process of stochastic scenario generation for cooling, heating, and power loads is shown in Fig. 2. The steps are as follows:

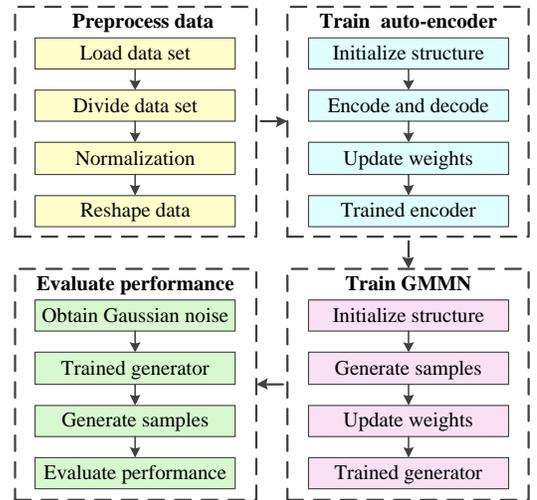

Fig. 2. Process of stochastic scenario generation.

1) Process data

The data set is divided into the training set and test set. 80% of the load curves are randomly selected to train the GMMN, and the remaining samples are used to evaluate the performance of the GMMN. Before feeding the load curves into the model, the load curves need to be normalized, otherwise the loss functions of the auto-encoder and the scenario generator may not converge. Therefore, the minimum-maximum normalization method is selected to transform load curves into the range of 0 to 1:

$$X_i^{'} = \frac{X_i - X_{\min}}{X_{\max} - X_{\min}} \quad (8)$$

where $X_i$ is the input data before normalization; $X_i^{'}$ is the input data after normalization; $X_{\max}$ the maximum value of the load; and $X_{\max}$ the minimum value of the load.

2) Train auto-encoder

After initializing the network structure, the normalized training samples are fed into the encoder. The decoder takes the low-dimensional latent variables output by the encoder as input data, and then outputs the reconstructed load curves. Next, real load curves and reconstructed load curves are used to calculate MSE and update the weights of the encoder and decoder. When the set number of iterations is reached, the encoder will be used in the training process of the scenario generator.

3) Train the GMMN

After initializing the network structure, the Gaussian noises are fed into the scenario generator to obtain new load curves similar to real load curves. The trained encoder transforms the real load curves and new load curves into latent variables for calculating the MMD loss function, which is used to update the weights of the scenario generator. When the set number of iterations is reached, the scenario generator will be used to generate stochastic scenarios for cooling, heating, and power loads.

4) Evaluate performance

After Gaussian noises are input into the trained scenario generator, the output result is de-normalized to obtain new load curves. Finally, the test set is used to measure whether the new load curves have similar temporal-spatial correlations and distribution characteristics with real load curves.

## IV. CASE STUDY

### A. Dataset and Simulation Tools

In order to fully verify the effectiveness of the algorithm proposed in this paper, the real dataset from the University of Texas at Austin is used for simulation and analysis [30], [31]. The Hal C. Weaver power plant and its associated facilities are in charge of providing all the cooling, heating, and power energies for the campus, which includes 70,000 students, staff, faculty, and 160 buildings. This dataset counts the hourly cooling, heating, and power needs from July 17, 2011 to September 4, 2012.

The programs of the GMMN for stochastic scenario generations of cooling, heating, and power loads are implemented in Spyder 3.2.8 with Tensorflow 1.12.0 and Keras 2.2.4 library. The programming language is the Python 3.7. Some parameters of the laptop are: 8 GB of memory, Intel(R) Core(TM) i5-10210U, the processor is @1.60GHz and 2.11GHz.

### B. Structure and parameters of the GMMN

In order to make the GMMN have high performance for stochastic scenario generations of cooling, heating, and power loads, the control variable method in [32] is utilized to search the suitable structures and parameters of the auto-encoder in the GMMN, as shown in Fig. 3.

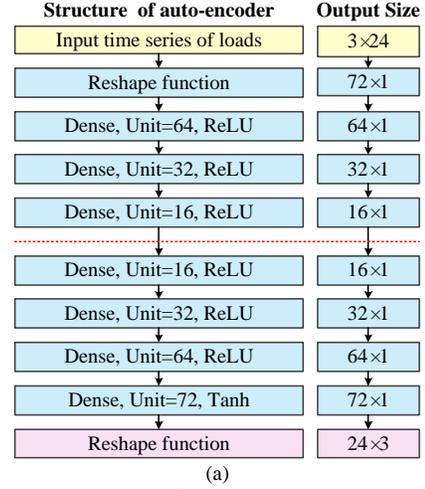

Fig. 3. Structure and parameters of the auto-encoder.

For the auto-encoder, its encoder consists of three dense layers with the rectified linear unit (ReLU) activation function. The number of neurons in dense layers is 64, 32, and 16, respectively. The structure of the decoder is similar to that of the encoder. It includes four dense layers with 16, 32, 64, and 72 neurons respectively. Except for the output layer, which uses the Tanh function as the activation function, the activation functions of other dense layers are ReLU functions. The optimizer is the Adam algorithm, and the maximum number for iterations is 500.

As shown in Fig. 4, the loss function of the auto-encoder is very close to 0 after 50 iterations. Therefore, the trained auto-encoder will be used to reduce the dimensions of real samples and generated samples.

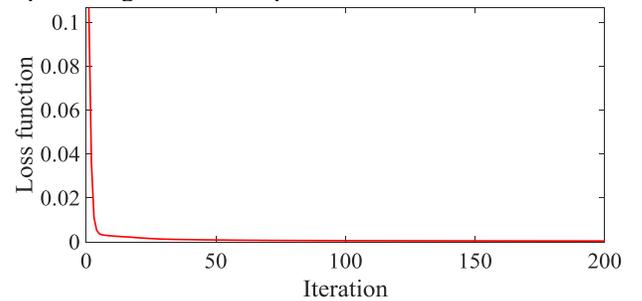

Fig. 4. Training evolution of the auto-encoder.

In order to find suitable hidden layers for the scenario generator, generators with different types and numbers of hidden layers are tested, and their mean loss functions are shown in Table I.

TABLE I
MENA LOSS FUNCTIONS IN DIFFERENT HIDDEN LAYERS

| Number of layers | Dense | LSTM | GRU | Conv2DTran |
|---|---|---|---|---|
| 1 | 0.48 | 0.53 | 0.45 | 0.49 |
| 2 | 0.48 | 0.51 | 0.44 | 0.46 |
| 3 | 0.47 | 0.47 | 0.55 | 0.40 |
| 4 | 0.46 | 0.48 | 0.54 | 0.43 |
| 5 | 0.46 | 0.46 | 0.49 | 0.48 |
| 6 | 0.52 | 0.55 | 0.54 | 0.51 |

Obviously, the loss functions of scenario generators constructed by density layer, long short-term memory (LSTM)

layer, and the gated recurrent unit (GRU) layer are larger than that of three transposed convolutional layers. Therefore, three transposed convolutional layers are used to construct the scenario generator for cooling, heating, and power loads, as shown in Fig. 5.

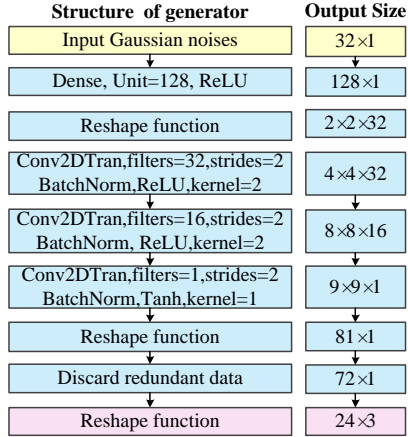

Fig. 5. Structure and parameters of the scenario generator.

For the scenario generator, it consists of one density layer and three transposed convolutional layers. Specifically, the number of neurons in the dense layer is 128 and the activation function is the ReLU function. The number of filters in transposed convolutional layers is 32, 16, and 1, respectively. Except for the last transposed convolutional layer, which uses the Tanh function as the activation function, the activation functions of other layers are ReLU functions. The optimizer is the Adam algorithm. Last but not least, the size of data generated by the output layer is 1×81, while the size of cooling, heating, and power loads is 1×72. Therefore, the last nine redundant data are discarded, and the first 72 data are used to calculate the MMD function.

After initializing structures of the scenario generator, a gradient descent method is needed to optimize the loss function. Fig. 6 shows the mean loss functions in many popular optimizers, such as SGD, RMSprop, Adadelta, Adagrad, Adam, Adamax, and Nadam.

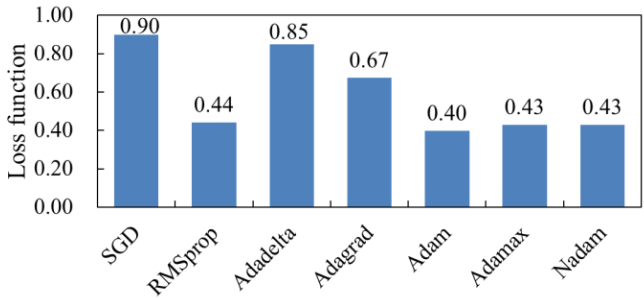

Fig. 6. Mean loss functions in different optimizers.

The GMMN has good performance when Nadam, Adamax, RMSprop, and Adam are used as optimizers. Specifically, the mean loss function of the Adam algorithm is slightly smaller than those of the first three algorithms. Therefore, the Adam algorithm is the most suitable optimizer for the GMMN in scenario generations of cooling, heating, and power loads.

Moreover, the learning rate is a configurable hyper-parameter that is used to control how quickly the GMMN is adapted to the scenario generation. To find a suitable learning rate for the GMMN, Fig. 7 shows the mean loss functions of different learning rates.

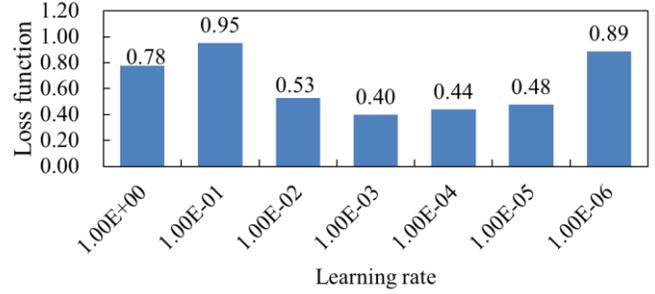

Fig. 7. Mean loss functions in different learning rates.

With the decrease of the learning rate, the loss function first decreases and then increases. The suitable learning rate of the GMMN is between 0.00001 and 0.001.

To find a suitable dimension for latent variables, auto-encoders with different dimensional latent variables are trained. Then, their encoders are used to train scenario generators, which produce a group of new samples. Finally, The MMD loss functions between real samples and new samples generated by scenario generators with different encoders are shown in Fig. 8.

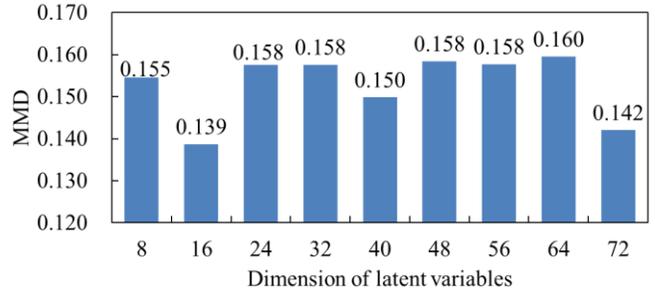

Fig. 8. MMD loss functions between real samples and new samples.

When the dimension of latent variables is equal to 16, the scenario generator has the strongest ability to capture the probability distribution of samples. 16 is a good starting point for dimensions of latent variables, and higher values or lower values may be fine for some data sets.

Normally, convergence speed is affected by the architecture and parameters of networks (e.g., hidden layers, loss functions, and optimizers). To observe the training stability and convergence performance of the designed GMMN, Fig. 9 intuitively compares the loss functions of the scenario generator in the GAN [15] and GMMN.

The MMD loss function of the GMMN decreases rapidly as the number of iterations increases. When the number of iterations is greater than 100, the MMD loss function of the scenario generator tends to a constant value, indicating that the GMMN has converged. Unlike GAN where the loss function fluctuates sharply and is difficult to converge, GMMN converges very quickly, and the entire training process is relatively stable.

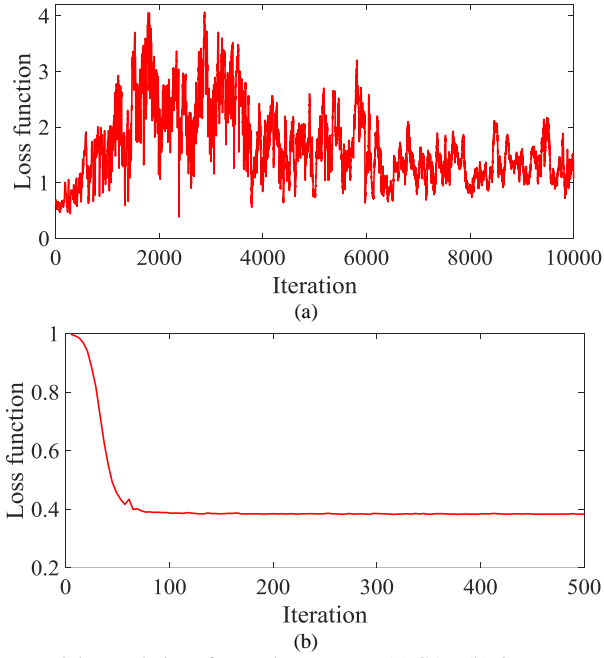

Fig. 9. Training evolution of scenario generator. (a) GAN (b) GMMN

*C. Simulation results and analysis*

To check whether the new samples generated by the GMMN and real samples have similar patterns, 2000 Gaussian noise samples are fed to the scenario generator of the GMMN to obtain the corresponding cooling, heating, and power curves. Then, a part of real samples are randomly selected from the test set, and the Euclidean distances between the new samples and the selected real samples are calculated. Finally, the selected real samples and their closest new samples are visualized, as shown in Fig. 10.

As shown in the first row of Fig. 10, the shapes of generated cooling, heating, and power curves are very similar to those of the real samples, so that it is hard to identify them. The GMMN accurately captures the hallmark characteristics of cooling, heating, and power load curves, such as large peaks, fast ramps, and fluctuation. Furthermore, the real samples of the test set did not participate in the training process of the GMMN, and the cooling, heating, and power load curves generated by the GMMN are consistent with the shapes of real samples in the test set, which shows that the GMMN has strong generalization ability.

In addition to shapes of multi-class load curves, some statistical properties between real samples and new samples should be verified. The auto-correlation function represents the temporal correlation of a time series, and capturing the correct temporal behavior is of great importance to operations of integrated energy systems. Therefore, the auto-correlation function is employed to compare the temporal correlation between real samples and new samples. Its mathematical formula is:

$$R(\tau) = \frac{E[(x_t - \mu)(x_{t+\tau} - \mu)]}{\sigma^2} \qquad (9)$$

where $x_t$ is a point of the load curve at the time $t$; $\mu$ is the mean of the load curve; $\sigma$ is the variance of the load curve; $\tau$ is the lag time; and $E$ is the expected value.

The second row of Fig. 10 shows the autocorrelation functions of the cooling, heating, and power load curves. It is found that the trends of autocorrelation functions of the generated samples closely resemble those of the real samples, which indicates that the GMMN is able to accurately capture the temporal correlation of the real cooling, heating, and power load curves.

The fluctuations and frequency-domain characteristics of cooling, heating, and power loads have a great influence on the operation of integrated energy systems. The power spectral density (PSD) represents the energy value of frequency components of load curves, and it is often utilized to measure the frequency-domain characteristics [33]. Its mathematical formula is:

$$P_{sd} = \lim_{T \to \infty} \frac{1}{T} \int_{-\infty}^{\infty} |x_T(t)|^2 \, dt \qquad (10)$$

where $P_{sd}$ is the power spectral density; and $T$ is the period. In this paper, the periodogram function from MATLAB2018a is employed to obtain the PSDs of these load curves.

The third row of Fig. 10 shows the PSDs of cooling, heating, and power load curves. It is obvious that the trends of PSDs between the generated samples for different loads and the real samples are basically the same, which shows that the real samples generated by the GMMN can reflect the fluctuation components of multi-class load curves at different frequencies of the real samples well.

Load duration curves represent the variation of a certain load in a downward form that the minimum value is plotted on the right and the maximum value is plotted on the left. The area under the load duration curves denotes the energy needs per day. Its mathematical formula is:

$$t_j = \sum_{i=1}^{m} q_{i,j}, q_{i,j} = \begin{cases} 1, P_i \geq P_j \\ 0, P_i < P_j \end{cases} \quad i = 1 \sim m, j = 1 \sim n \qquad (11)$$

where $m$ is the size of the load curve; $n$ is the number of intervals for load curves; and $t_j$ is the time when the loads are greater than $j^{th}$ element $P_j$ of the load curve.

From the fourth row of the Fig. 10, it can be found that the load duration curves of the real samples and the generated samples are extremely similar, and the areas enclosed by the X-axis and Y-axis are basically the same, which shows that the total energy consumption of the cooling, heating, and power load curves generated by the GMMN in one day is consistent with the actual scenarios.

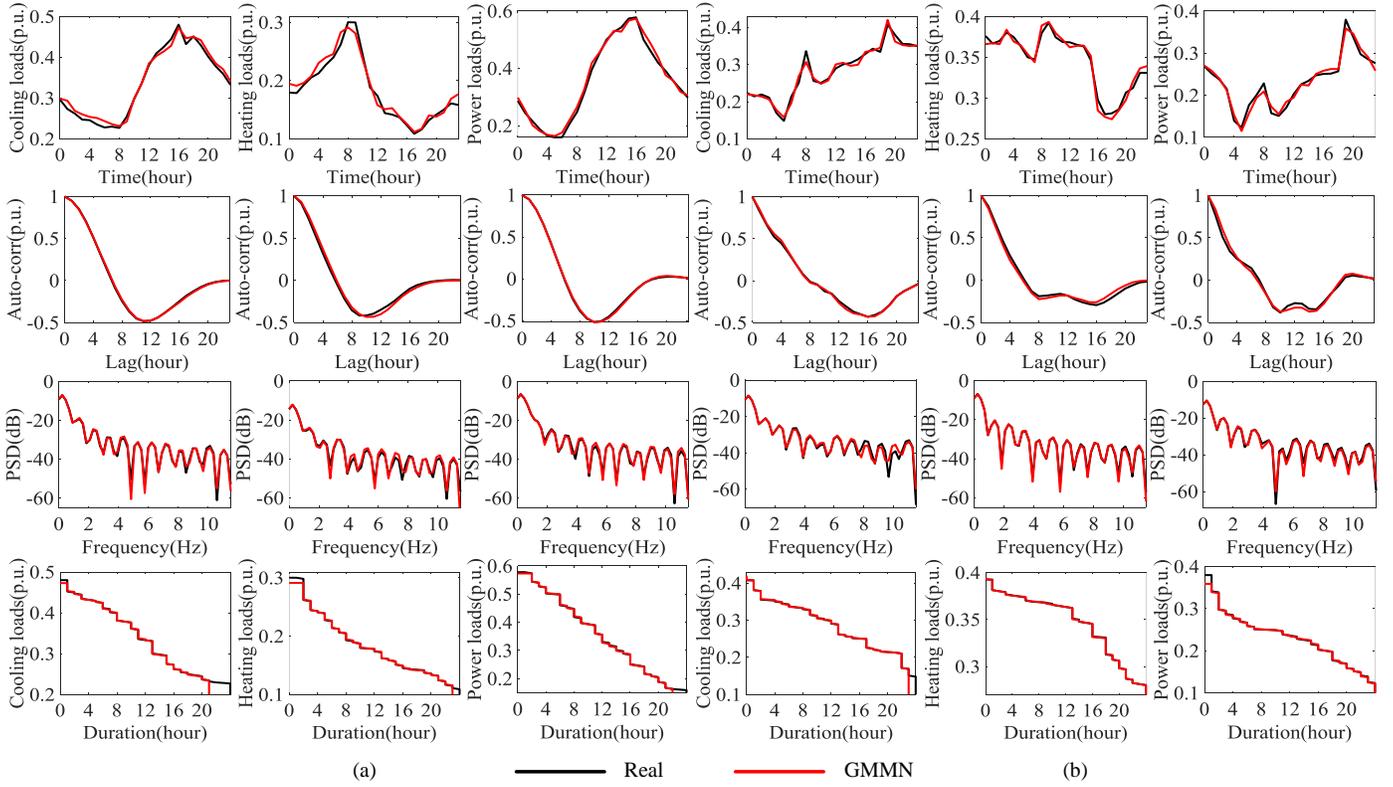

(a) ——— Real ——— GMMN (b)

Fig. 10. Visualization of real samples from the test set and new samples generated by the GMMN. (a) Sample 1. (b) Sample2.

As one of the best indicators measuring the association between continuous variables, the Pearson correlation coefficient is often used to evaluate the linear relationship of load curves at various look-ahead times [4]. Its mathematical formula is:

$$p_{xy} = \frac{\sum_{i=1}^{m}(x_i - \bar{x})(y_i - \bar{y})}{\sqrt{\sum_{i=1}^{m}(x_i - \bar{x})^2}\sqrt{\sum_{i=1}^{m}(y_i - \bar{y})^2}} \quad (12)$$

where $p_{xy}$ is the Pearson correlation coefficient between $x$ and $y$; $\bar{x}$ is the mean of $x$; and $\bar{y}$ is the mean of $y$.

To validate whether new samples generated from the GMMN have a similar temporal correlation as the real samples, Fig. 11, Fig. 12, and Fig. 13 visualize the covariance matrix of real samples and generated samples for cooling, heating, and power load curves.

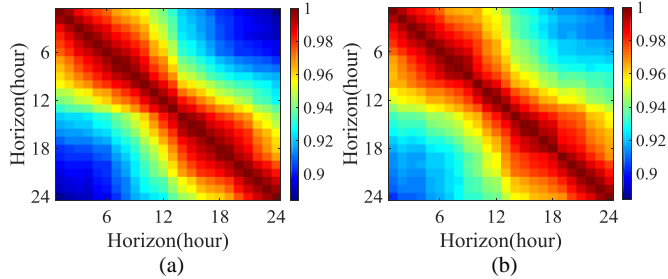

Fig. 11. The Pearson correlation matrix of cooling loads. (a) Real samples. (b) New samples generated by the GMMN.

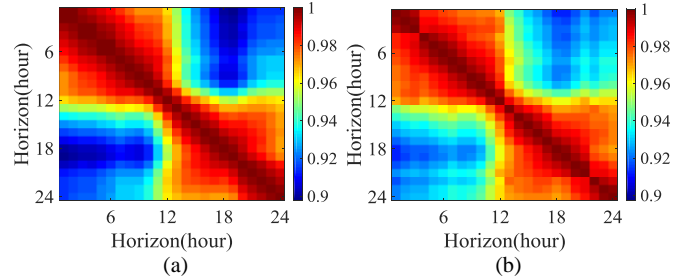

Fig. 12. The Pearson correlation matrix of heating loads. (a) Real samples. (b) New samples generated by the GMMN.

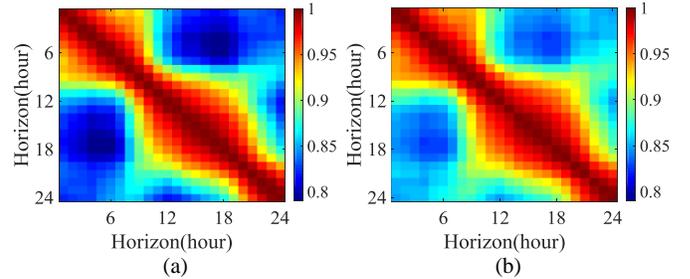

Fig. 13. The Pearson correlation matrix of power loads. (a) Real samples. (b) New samples generated by the GMMN.

The following conclusions can be drawn from the Fig. 11 to Fig. 13: 1) Although the Pearson correlation coefficients between current loads and previous loads decrease with the increase of time, they are always greater than 0.8, which indicates that there is a strong temporal correlation in cooling, heating, and power load curves. Specifically, the temporal correlation in heating load curves is the strongest, since the Pearson correlation coefficients between current heating loads and previous heating loads are always greater than 0.9, while the temporal correlation in power load curves is the weakest. 2) The covariance matrix element values of real samples and generated samples are very similar, which indicates that GMMN is able to accurately capture the temporal dependency of cooling, heating, and power load curves without any model assumptions being made during the training process.

Previous works have shown that there are strong spatial correlations between the cooling, heating, and power loads, which have a significant impact on the operation and planning of integrated energy systems [1], [2]. Therefore, it is necessary to account for the spatial correlations when generating stochastic scenarios.

Specifically, from the two samples in the first row of Fig. 10, it can be seen qualitatively that the cooling loads and power loads are positively correlated, while cooling loads and power loads are negatively correlated with heating loads. GMMN takes into account the spatial correlation among multi-class loads when generating the new stochastic scenarios, which is in line with the actual scenarios.

Moreover, the spatial correlations among cooling, heating, and power loads are quantitatively analyzed using calculating the Pearson correlation coefficients, as shown in Fig. 14.

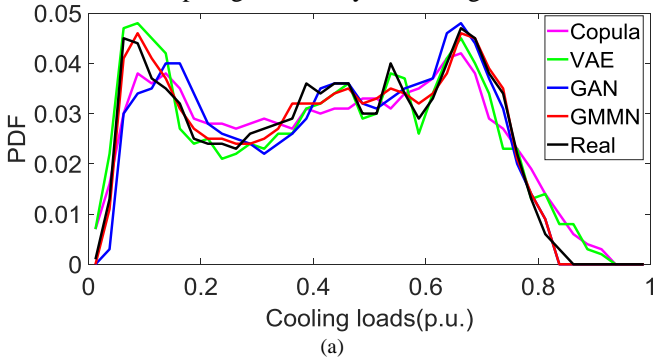

Fig. 14. The Pearson correlation matrix among multi-class loads. (a) Real samples. (b) New samples generated by the GMMN.

Obviously, the Pearson coefficient matrix of new samples has a small difference from that of the real sample, and the maximum error is 0.029, which indicates that the GMMN can well capture the spatial correlation among multi-class loads.

Besides verifying the above properties, Fig. 15 shows the probability distribution functions (PDFs) of historical samples and new samples generated by the GMMN and popular baselines such as the Copula method [7], VAE [12], and GAN [14]. In addition, Table II quantitatively calculates the Euclidean distances between the PDFs of real samples and the PDFs of new samples generated by different generative models.

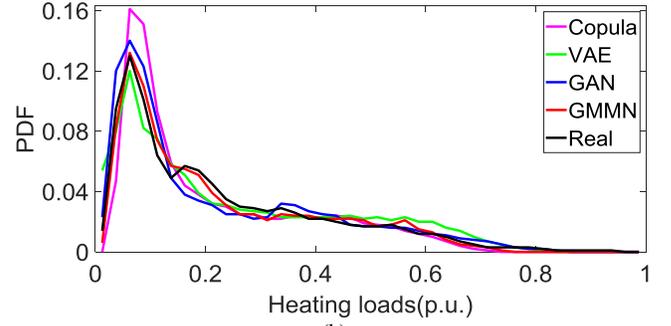

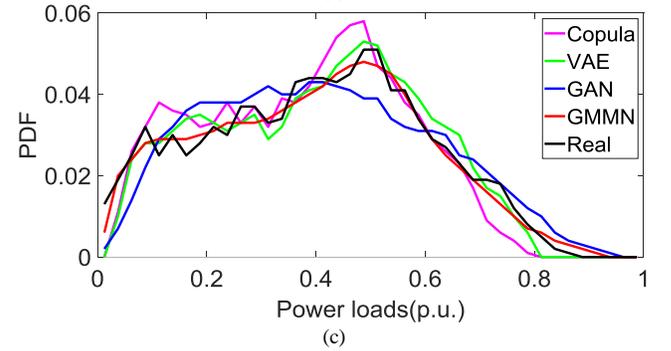

Fig. 15. PDFs of multi-class loads. (a) Cooling loads. (b) Heating loads. (c) Power loads.

TABLE II
EUCLIDEAN DISTANCES BETWEEN PROBABILITY DENSITY FUNCTIONS

| Mehtod | Cooling load | Heating load | Power load |
|---|---|---|---|
| Copula | 0.033 | 0.088 | 0.038 |
| VAE | 0.027 | 0.059 | 0.027 |
| GAN | 0.030 | 0.055 | 0.030 |
| GMMN | 0.013 | 0.027 | 0.017 |

It is found that the traditional Copula method has the worst ability to capture the probability distribution characteristics of cooling, heating, and power load. The performance of the VAE and the GAN is close. Moreover, the difference of PDFs between real samples and the new samples generated by the GMMN is very small, and three PDFs of the GMMN are closer to those of real samples than the existing methods (e.g., Copula method, VAE, and GAN), which indicates the capability of the GMMN to generate new samples for cooling, heating, and power loads with the correct marginal distributions.

## V. DISCUSSION

In this paper, a new data-driven approach is proposed to generate stochastic scenarios for cooling, heating, and power loads. The performance of the proposed approach has been compared with popular baselines on a dataset from the University of Texas at Austin. The simulation results show that the GMMN has a better performance than VAE, GAN, and the Copula method. However, the proposed method cannot generate labeled samples directly, because it belongs to unsupervised generative networks. Therefore, GMMN may be extended to a supervised generative network to generate conditional load curves (e.g., heavy loads or light load) by employing the idea of conditional GAN [34] or conditional VAE [35].

In the field of computer visions, GMMN has showed outstanding performance for many high-dimensional image datasets, such as MNIST handwritten digit database, Yale Face Database, and Toronto Face Dataset. Therefore, GMMN should have the potential for scenario generations of the large-scale integrated energy systems with a large number of loads. Moreover, structure and parameters of the GMMN in this paper should be fine-tuned to accommodate scenario generations for large-scale loads. Some publications on how to adjust the structures of deep neural networks can be found in [4], [32].

Last but not least, the applications of the GMMN are not limited to scenario generation of loads. It may be generalized to other data generation tasks of integrated energy systems, such as scenario prediction of renewable energy sources.

## VI. Conclusion And Future Works

To improve the quality of stochastic scenario generation for cooling, heating, and power loads, this paper proposes a novel data-driven method based on the GMMN. Through the simulation analysis on a real dataset, the following conclusions are obtained:

1) Unlike the GAN where the loss function fluctuates sharply and is difficult to converge, GMMN converges very quickly, and the entire training process is relatively stable. The suitable learning rate of the GMMN is between 0.00001 and 0.001. The Adam algorithm is the most suitable optimizer for the GMMN in scenario generations of cooling, heating, and power loads. Besides, GMMN fits the probability distribution characteristics of cooling, heating, and power loads more than some popular methods such as the Copula method, VAE, and GAN.

2) Simulation results show that the GMMN accurately captures the hallmark characteristics (e.g., large peaks, fast ramps, and fluctuation), frequency-domain characteristics, and temporal correlation of cooling, heating, and power load curves. In addition, the energy consumption of generated samples closely resembles that of real samples.

3) GMMN takes into account the spatial correlation among multi-class loads when generating the new stochastic scenarios, which is in line with the actual scenes.

## Acknowledgment

This manuscript was supported by the China Scholarship Council. The authors are very grateful for their help.